\documentclass[aps, pre, floatfix,
twocolumn
]{revtex4-1}

\usepackage{graphicx}
\usepackage{bm}
\usepackage{subfigure}
\usepackage{hyperref}
\usepackage{amsmath}
\usepackage{color}

\newcommand{\red}[1]{{\color{black} #1}}

\begin{document}

\title{Kinetics of random sequential adsorption of two-dimensional shapes on~a~one-dimensional line}

\author{Micha\l{} Cie\'sla$^1$}
  \email{michal.ciesla@uj.edu.pl}
\author{Konrad Kozubek$^1$}
  \email{konrad.p.kozubek@gmail.com}
\author{Piotr Kubala$^1$}
  \email{pkua.log@gmail.com}
 \author{Adrian Baule$^2$}
  \email{a.baule@qmul.ac.uk}
  
  \affiliation{$^1$M.\ Smoluchowski Institute of Physics, Department of Statistical Physics, Jagiellonian University, \L{}ojasiewicza 11, 30-348 Krak\'ow, Poland\\
  $^2$School of Mathematical Sciences, Queen Mary University of London, London E1 4NS, United Kingdom}


\date{\today}

\begin{abstract}
Saturated random sequential adsorption packings built of two-dimensional ellipses, spherocylinders, rectangles, and dimers placed on a one-dimensional line are studied to check analytical prediction concerning packing growth kinetics [A. Baule, Phys. Rev. Let. 119, 028003 (2017)]. The results show that the kinetics is governed by the power-law with the exponent $d=1.5$ and $2.0$ for packings built of ellipses and rectangles, respectively, which is consistent with analytical predictions. However, for spherocylinders and dimers of moderate width-to-height ratio, a transition between these two values is observed. We argue that this transition is a finite size effect that arises for spherocylinders due to the properties of the contact function. In general, it appears that the kinetics of packing growth can depend on packing size even for very large packings.

\end{abstract}

%
\maketitle
\section{Introduction}
Random sequential adsorption (RSA) \cite{Evans1993} is a model of random packing generation in which objects are added to the packing according to the following scheme:   
\begin{itemize}
    \item a virtual object's position and orientation are selected randomly inside a packing;
    \item if the object does not intersect with previously added particles, it is added to the packing and holds its position and orientation unchanged;
    \item if the object intersects with any of the existing objects, it is removed and abandoned.
\end{itemize}
These iterations are repeated until the packing becomes saturated, which means that there is no possibility of placing another object there.
RSA owes its popularity to the observation that such packings resemble monolayers obtained in irreversible adsorption processes \cite{Feder1980,Onoda1986}. From the theoretical point of view, RSA packings are interesting as probably the simplest, yet not trivial random packing model which accounts for excluded volume effects. In contrast to more popular random close packings, where neighboring particles are in touch, the RSA packings have well-defined mean packing fraction, which is an additional asset for numerical and theoretical studies \cite{Torquato2000, Torquato2010}.  
\red{However, only for some specific two-dimensional shapes, there exist algoritms, which generates saturated RSA packings \cite{Zhang2013, Zhang2018, Haiduk2018, Kasperek2018, Ciesla2019, Ciesla2020} and estimation of the mean saturated packing fraction is straightforward. In general case, the knowledge about packing growth kinetics is needed because above described RSA protocol does not give any hint when packing become saturated and no other particle can be added to it. Therefore, typically the packing generation is interrupted after some finite number of iterations and the number of particles in saturated state is estimated using } the power-law:
\begin{equation}
    \theta(t) = \theta - A t^{-1/d}.
    \label{eq:fl}
\end{equation}
Here $\theta(t)$ and $\theta$ are the mean packing fraction after $t$ iterations and at saturation, respectively and $A$ is a positive constant \cite{Feder1980}. Parameter $d$, for packings built of spherically symmetric particles, is equal to the packing dimension \cite{Pomeau1980, Swendsen1981}. For two-dimensional packings built of anisotropic shapes, it is typically equal to $3$ \cite{vigil1989, Viot1992, Shelke2007, Ciesla2015, Ciesla2016}, and therefore it was assumed that it is equal to the number of shape's degrees of freedom \cite{Hinrichsen1986,Ciesla2013}. Situation changes when two-dimensional shapes are placed on a one-dimensional line. Recently, Baule provided analytical arguments that the RSA packing built of ellipses, whose centers are on a one-dimensional line, grows faster than similar packings built of rectangles or spherocylinders \cite{Baule2017}, which is different than for two-dimensional packings \cite{vigil1989,Viot1992,Ciesla2016}. The difference in growth kinetics originates in the properties of the contact function, which is defined as the separation distance at which two particles of given orientations are in contact. For ellipses the contact function is always analytical, but it can be non-analytical, i.e. piecewise-continuous, for rectangles and spherocylinders depending on the orientations of the particles. 

The main aim of this study is to check this effect numerically, using recent algorithms that allow generating strictly saturated RSA packings \cite{Haiduk2018,Kasperek2018}. Besides ellipses, rectangles and spherocylinders, packings built of two-dimensional dimers are also analyzed to study if there is any difference for non-convex shapes. Additionally, the dependence of the packing fraction on the anisotropy of particles that build the packing and the scaling of the number of RSA iterations needed to generate a saturated packing with the size of the packing are studied. For spherically symmetric particles this scaling is governed by the same parameter as the kinetics of packing growth \cite{Ciesla2017}.
\section{Model}
\begin{figure}[htb]
  \vspace{0.5in}
  \includegraphics[width = 0.8\columnwidth]{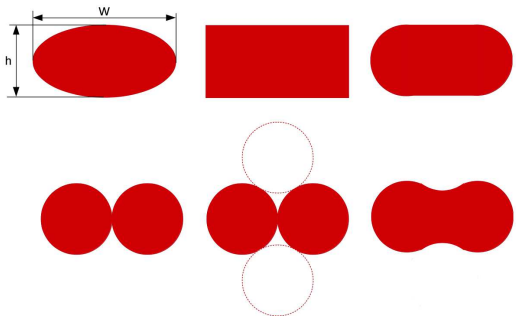}
  \caption{Four types of shapes used for RSA packing generation. All the shapes were characterized by their anisotropy $x$ which is defined as width-to-height ratio. Note that dimer, which is built of two identical disks (left bottom panel) is equivalent to the smooth shape in the right bottom panel.}
  \label{fig:model}
\end{figure}

Random packings of ellipses, rectangles, spherocylinders and dimers (see Fig.~\ref{fig:model}) were generated using algorithms for saturated packing generation \cite{Haiduk2018,Kasperek2018,Ciesla2020}. \red{These algorithms trace regions where subsequent shapes can be added. Note, that each figure placed on a line blocks some area around it because placing there the center of the next shape will cause intersection. The size of this area varies with the orientation of the next figure. The algorithms trace these regions and when they fill the whole line for any orientation of the shape that can be added to the packing then the packing is saturated. This approach was firstly used for generation of saturated RSA packings built of spherically symmetric figures \cite{Wang1994, Ebeida2012, Zhang2013} and further was extended to some anisotropic shapes like ellipses and spherocylinders \cite{Haiduk2018} rectangles \cite{Kasperek2018} and dimers \cite{Ciesla2020}.} 
Centers of shapes were placed on a one-dimensional line segment. Typically, the length of this line segment was  $L=10^6$, but for spherocylinders and dimers different sizes: $L \in [10^2, 10^7]$ were also used. Periodic boundary conditions were used to minimize finite-size effects \cite{Ciesla2018}. Each shape had a unit surface area. It is worth noting that this assumption is one of many other possibilities of comparing results obtained for shapes with different anisotropies. For example, in the study of Chaikin et al., different ellipses had the same length of short axis \cite{Chaikin2006}. Simulations were performed for width-to-height ratios $x<3$ for ellipses, rectangles and spherocylinders. For dimers the highest studied anisotropy is $x=2.4$. Here,  considering their smooth version, it ceases to be connected for $x>1+\sqrt{3}$. However, the equivalence between two disks and smoothed dimers breaks already for $x \ge 1+\sqrt{2}$, because there it becomes possible to arrange non-intersecting disks in a way that they correspond to intersecting dimers.  For each particular shape and the packing size $L$, $100$ independent saturated random packings were generated and analyzed to determine the mean saturated packing fraction and the kinetics of packing growth. In particular:
\begin{eqnarray}
\label{theta}
    \theta & =  &\frac{1}{100}\sum_{i=1}^{100} \theta_i \nonumber \\
    \sigma(\theta) & = & \frac{1}{100} \left[ \sum_{i=1}^{100} (\theta - \theta_i)^2 \right]^\frac{1}{2},
\end{eqnarray}
where $\theta_i$ is the coverage of $i$-th packing, $\theta$ is the mean coverage and $\sigma(\theta)$ is its standard deviation that estimates its error. \red{The number of packings used in these calculations guarantees that the statistical error of studied properties will be negligible.} To compare results for differently sized packings, the number of iterations $n$ was measured using dimensionless time units
\begin{equation}
\label{t}
    t = \frac{n}{L}.
\end{equation}
\section{Results and discussion}
\begin{figure}[htb]
  \vspace{0.5in}
  \includegraphics[width = 0.8\columnwidth]{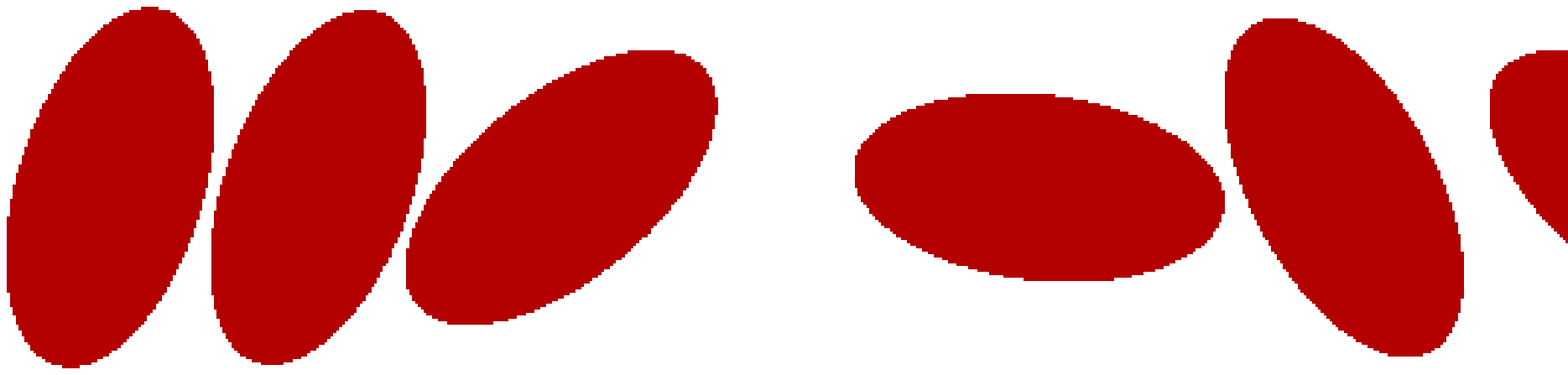}
  \includegraphics[width = 0.8\columnwidth]{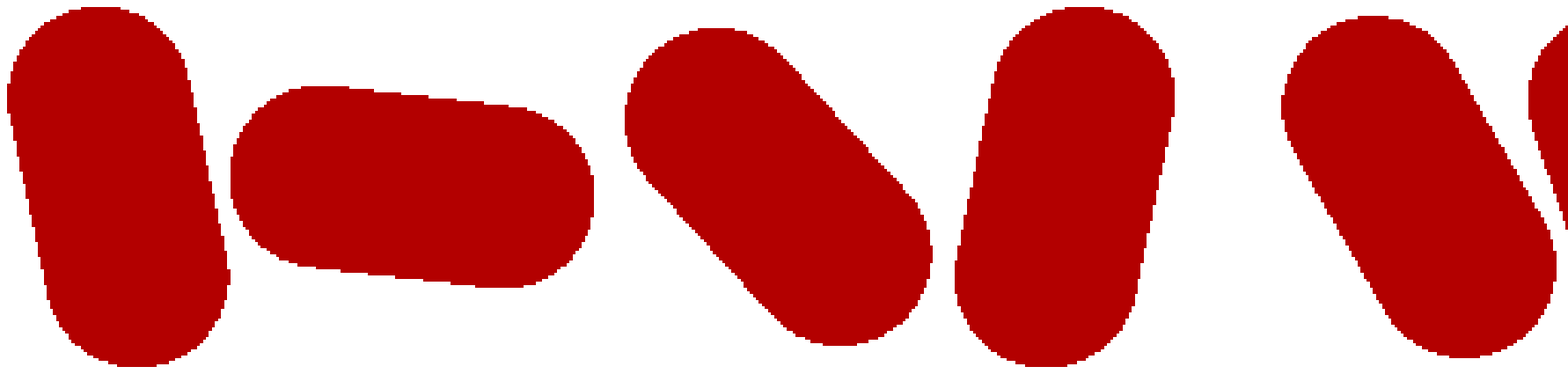}
  \includegraphics[width = 0.8\columnwidth]{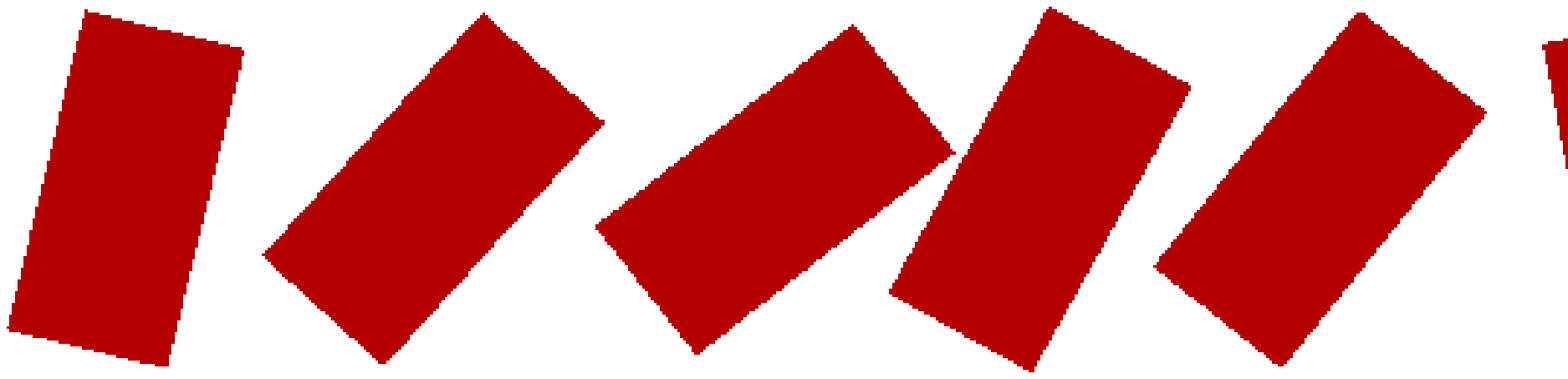}
  \includegraphics[width = 0.8\columnwidth]{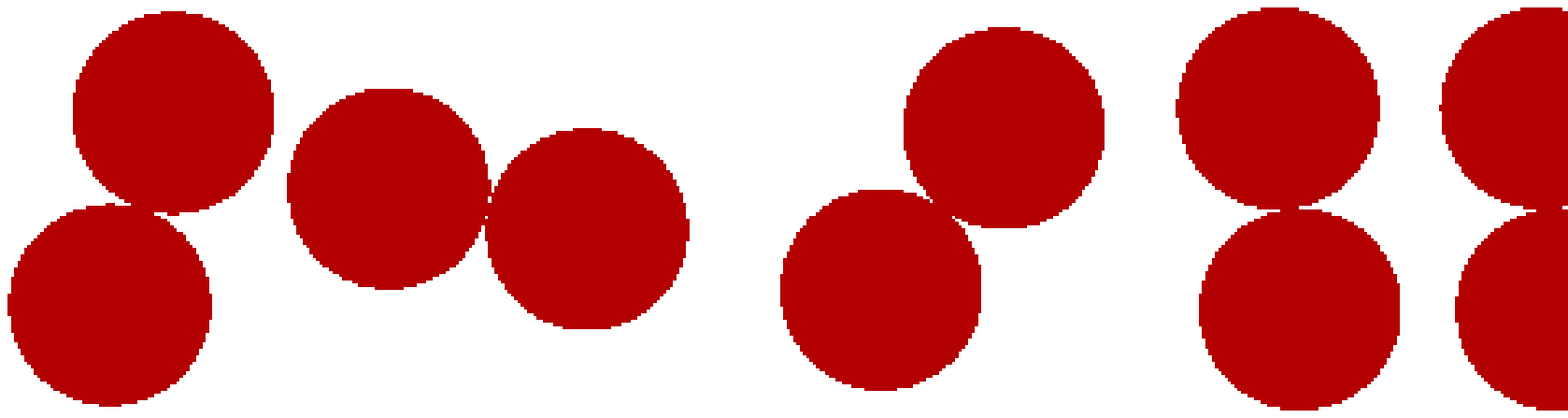}
  \caption{Fragments of illustrative saturated random packings of ellipses, spherocylinders, rectangles and dimers. The width-to-height ratio for all of these shapes is $x=2.0$.}
  \label{fig:examples}
\end{figure}

Fragments of illustrative packings are shown in Fig. \ref{fig:examples}.
\subsection{Mean saturated packing fraction}
\begin{figure}[htb]
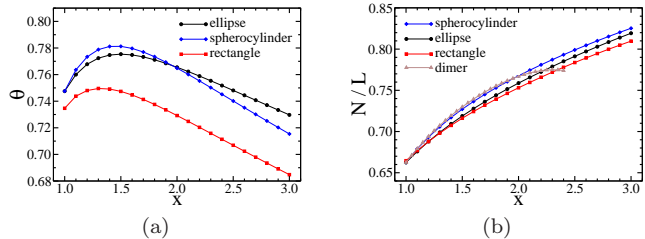

  \vspace{0.5in}
  \subfigure[]{\includegraphics[width = 0.45\columnwidth]{q_x}}
  \hspace{0.05\columnwidth}
  \subfigure[]{\includegraphics[width = 0.45\columnwidth]{q_x_n}}
  \caption{The dependence of the mean coverage ratio $\theta$ (a) and the mean density of shapes (b) on the width-to-height ratio $x$ for studied shapes. Dots are numerical data obtained for packing size $L=10^6$, and solid lines are to guide the eye.}
  \label{fig:q_x}
\end{figure}

In contrast to the case where shape and packing dimensions are the same, here the mean saturated packing fraction can be defined at least in two ways: using the coverage ratio $\theta$, or using the mean density of shapes $N/L$. Because a single, anisotropic object covers a different amount of a line depending on its orientation, these definitions can lead to different results, and, what is even more interesting, to different conclusions -- see Fig. \ref{fig:q_x}.  
For $x=1$, in case of ellipses and spherocylinders the coverage ratio should be equal to R\'enyi car parking constant $\theta = 0.7475979...$ \cite{Renyi1958}. Here, numerical simulation gives $\theta(1.0) = 0.747573 \pm 0.000022$, which agrees with theoretical predictions within slightly more than one standard deviation error range. For squares this value is a little smaller $\theta(1.0) = 0.734679 \pm 0.000021$, which seems counter-intuitive. In contrast to disks, the case of squares corresponds to packing of variable size segments on a one-dimensional line, and such shapes typically form denser packings \cite{Subashiev2007, Barbasz2013}. However here, the space for placing another object can be blocked due to crossing in the additional (in this case the second) dimension, which was not the case in already studied RSA of multidispersive shapes. For growing anisotropy the coverage ratio increases and reaches its maximum for moderate width-to-height ratio $x$. This is typical for a packing built of anisotropic shapes \cite{vigil1989, Viot1992, Ciesla2015, Ciesla2016}. Here, the highest observed coverage ratios are $\theta(1.5) = 0.775380 \pm 0.000019$ for ellipses, $\theta(1.5) = 0.781249 \pm	0.000020$ for spherocylinders and $\theta(1.3) = 0.749575 \pm 0.000016$ for rectangles. Results for dimers were not included in this discussion, because of the equivalence of different shapes (see Fig. \ref{fig:model}), which results in different values of packing fractions. 

Packing density behaves differently. It starts from a lower value than the coverage ratio, which is a consequence of the normalization used. Because of the unit surface area, the diagonal of the disk is $2 / \sqrt{\pi} \approx 1.128...$. Therefore, packing densities for disks are smaller by this factor than the coverage ratio, which does not depend on the size of the shape. It is worth noting that for $x=1$, packing densities for all studied shapes are almost equal to each other. With an increase of the width-to-height ratio, the objects' density grows monotonically, which is a consequence of the assumption that all shapes have the same surface area. Therefore, for larger anisotropy $x$, the shapes become thinner and the expected value of their cross-section with the line becomes lower. Thus, more of them can be placed there. However, this reasoning does not work in case of dimers, because their height does not decrease as fast as for other studied shapes. Therefore, here we observe the maximum density of $0.774962 \pm 0.000027$ for $x=2.3$.

\red{It is worth noting that the mean packing fraction can also be determined by studying cumulative distribution function:
\begin{equation}
\label{cdf}
    \mathrm{CDF}(\theta) = Prob(\theta_i < \theta),
\end{equation}
where $\theta_i$ is the packing fraction of the $i$-th random packing. For infinitely large packings the $\mathrm{CDF}(\theta)$ is a step function, but for finite ones it grows continuously from $0$ t o$1$ -- see Fig. \ref{fig:cdf}. 
\begin{figure}[htb]
  \vspace{0.5in}
  \includegraphics[width = 0.7\columnwidth]{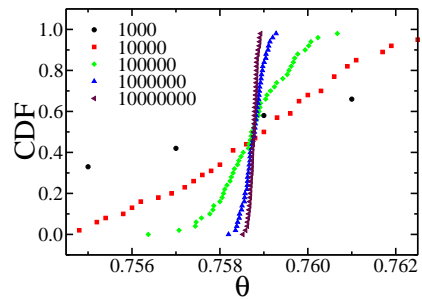}
  \caption{The cumulative distribution function (\ref{cdf}) for packing of different length. Dots corresponds to numerical data obtained by studying $100$ independent random packings.}
  \label{fig:cdf}
\end{figure}
The mean packing fraction at the limit of infinite packing can be estimated by finding the crossing of the CDF's for different packing sizes. This method is especially useful for studying RSA on lattices \cite{Vandewalle2000, Buchini2019, Ramirez2019, Ramirez2019a, Ramirez-Pastor2019, Pasinetti2019}. However, in our case, the precision given by \ref{theta} is enough due to quite large size of packings used in this study.}
\subsection{The kinetics of packing growth}
Although in general the kinetics of packing fraction growth and particles density growth is different, here, to be consistent with the previous theoretical study \cite{Baule2017}, we will focus on the second one. We have also checked that for large enough $t$, both kinetics converge to each other; thus, the presented results should be universal.

\begin{figure}[htb]
  \vspace{0.5in}
  \includegraphics[width = 0.7\columnwidth]{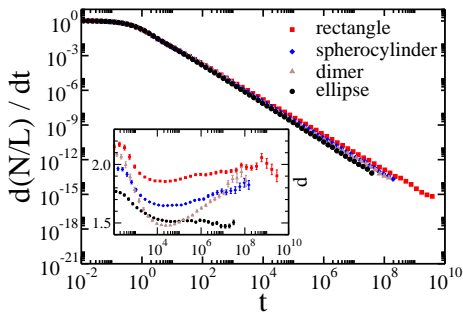}
  \caption{The dependence of the increments of the mean particles density on time for ellipses, spherocylinders, rectangles and dimers of width-to-height ratio $x=2.0$. The inset shows the dependence of the exponent $d$ from Eq. \ref{eq:fl} on the dimensionless time $t$ \red{(\ref{t})}. The value of parameter $d$ for a given time $t$ was estimated as a best fit of Eq. \ref{eq:fl} to numerical data in the range $[10^{-2}t, t]$. Ends of the lines correspond to the time $t_{min}$ for which the first of $100$ generated packing saturates. }
  \label{fig:d_t}
\end{figure}

\begin{figure}[htb]
  \vspace{0.5in}
  \includegraphics[width = 0.7\columnwidth]{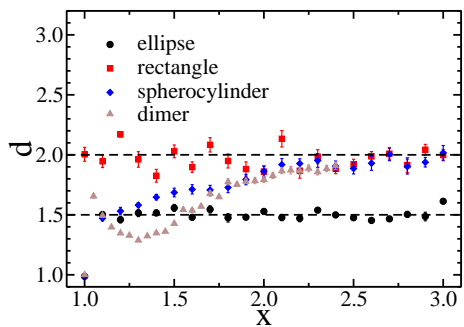}
  \caption{The dependence of the exponent $d$ near saturation on the width-to-height ratio $x$ for ellipses, spherocylinders and rectangles. Dashed lines correspond to $d=1.5$ and $d=2$ derived analytically for shapes with analytical and non-analytical contact function \cite{Baule2017}.}
  \label{fig:d_x}
\end{figure}

Examples of kinetics of the mean density of particles in the packing are shown in Fig. \ref{fig:d_t}.
Besides the fact, that the kinetics in a log-log scale for high enough value of $t$ seem to agree with the power-law (\ref{eq:fl}) for all shapes, the more detailed analysis shows that the slopes of these lines are apparently not constant -- see the inset in Fig. \ref{fig:d_t}. Here we are mainly interested in the asymptotic value of the parameter $d$, however, the accuracy of the power-law (\ref{eq:fl}) fitting decreases near saturation due to poor statistics -- there are only very few shapes added to the packing there, so a single placing event can significantly affect the result. Additionally, the number of iterations after a packing becomes saturated is a random variable described by heavy-tail probability density function \cite{Ciesla2017}, so after the same number of iterations, different packings are not similarly close to saturation. Therefore, as a final value of the parameter $d$ the result of fitting in the range $[10^{-3}t_\text{min}, 10^{-1}t_\text{min}]$ was used, where $t_\text{min}$ is the smallest observed number of iterations needed to generate saturated packing for the given shape. The dependence of such exponent $d$ on the anisotropy of the shape packed is shown in Fig.~\ref{fig:d_x}.
The results for ellipses and rectangles are in a good agreement with analytical predictions~\cite{Baule2017}. Interesting behavior is observed for spherocylinders and dimers. For spherocylinders of small anisotropy ($x \le 1.3$) the RSA kinetics is the same as for ellipses, but for large ones ($x>2.0$) it resembles kinetics of packings built of rectangles. For medium anisotropies, a continuous transition between these two limits is observed. A similar smooth transition is observed for dimers.

\begin{figure}[htb]
  \vspace{0.5in}
  \includegraphics[width = 0.7\columnwidth]{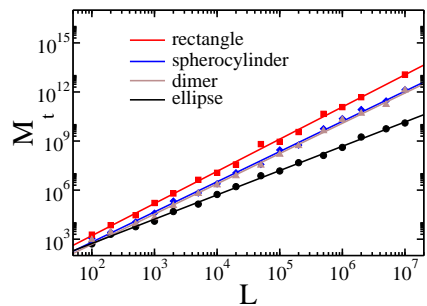}
  \caption{The dependence of the median of the number of iterations needed to generate saturated packing on packing size. Dots correspond to numerical data for shapes of anisotropy $x=2.0$ and solid lines are power fits corresponding to $d = 1.486$, $d=1.851$, $d = 1.852$ and $d = 1.967$ for ellipses, dimers, spherocylinders and rectangles, respectively.}
  \label{fig:median}
\end{figure}

\begin{figure}[htb] 
  \subfigure[]{\includegraphics[width = 0.7\columnwidth]{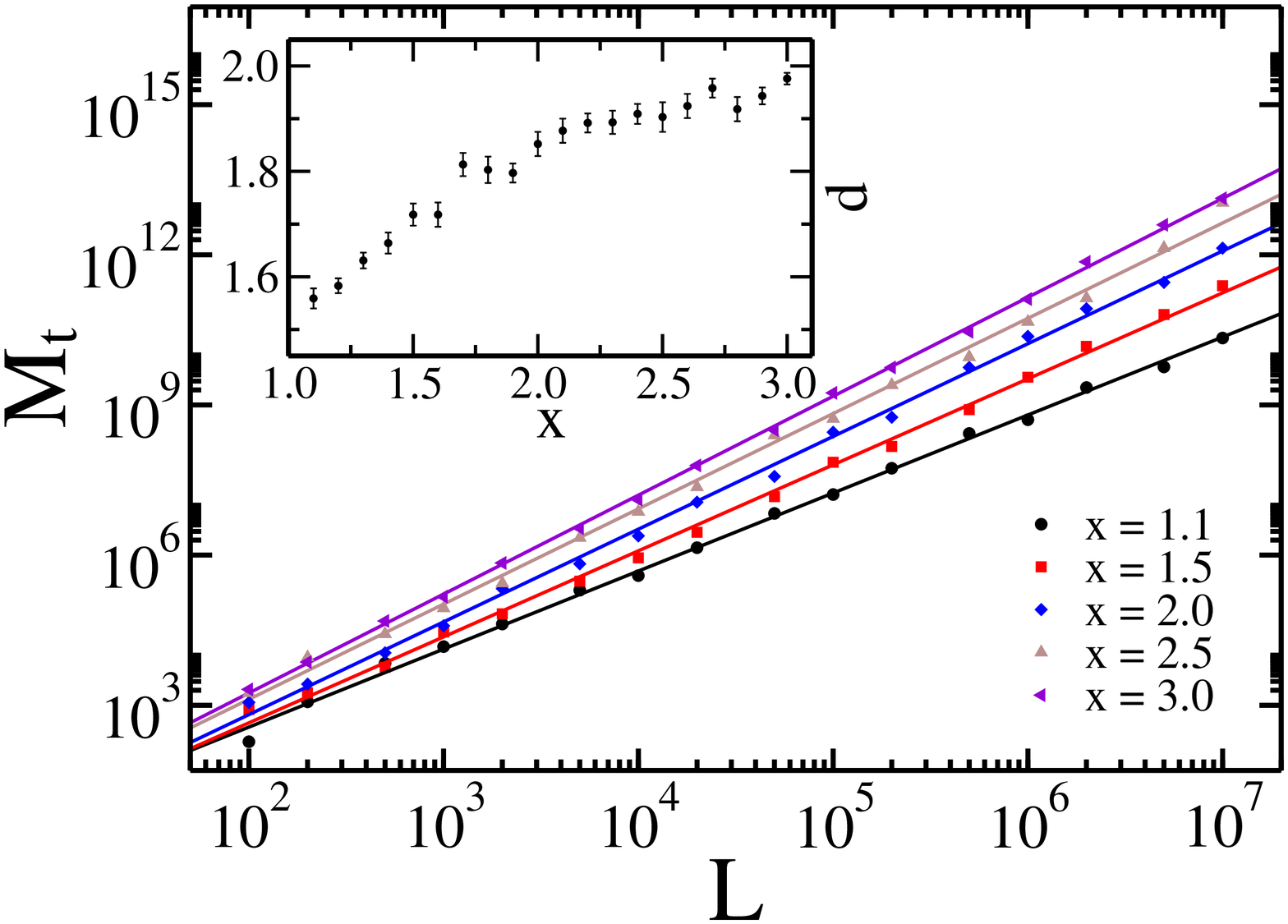}}
  \subfigure[]{\includegraphics[width = 0.7\columnwidth]{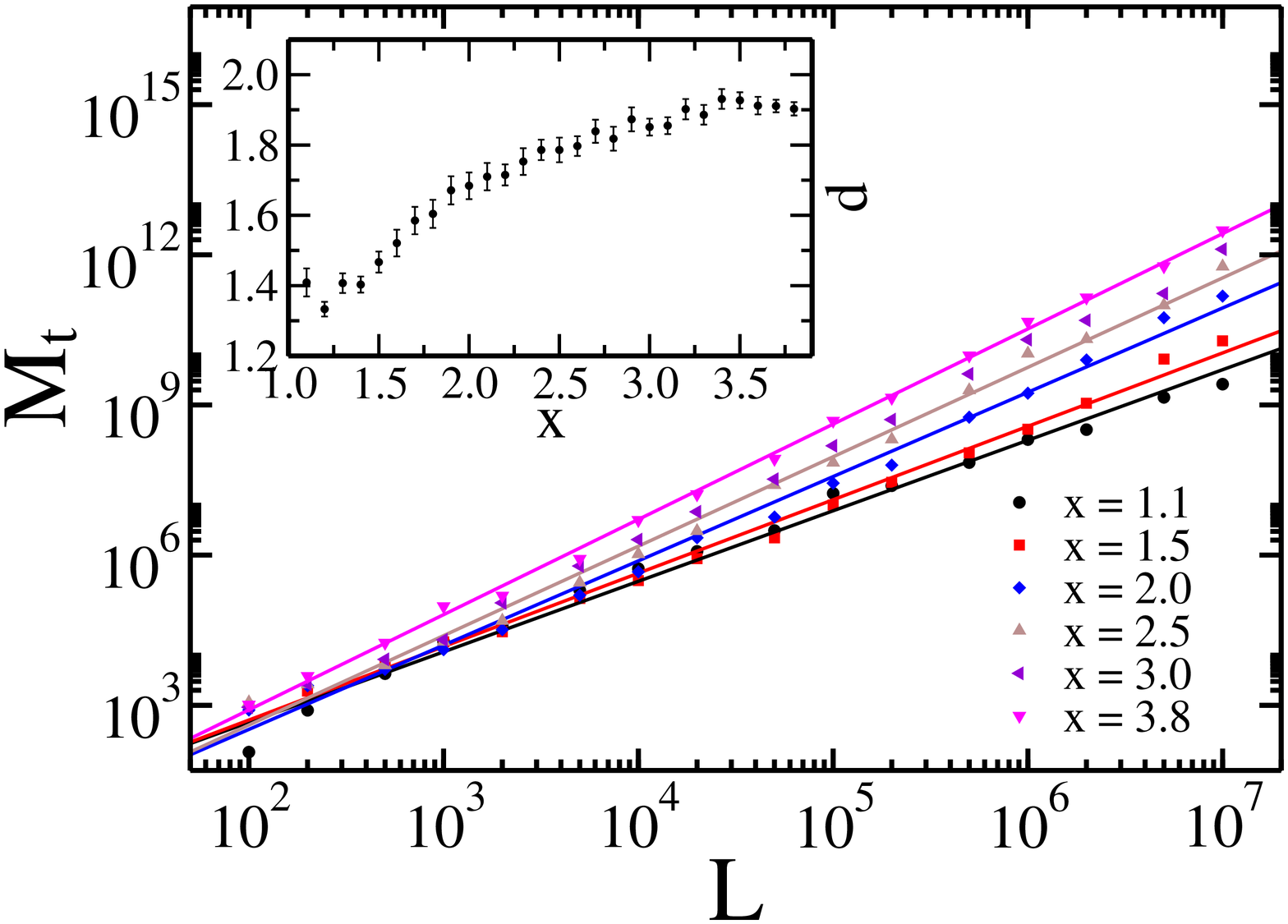}}
  \caption{The dependence of the median of the number of iterations needed to generate saturated packing on packing size for spherocylinders (a) and dimers (b). Insets show the dependence of parameter $d$ determined from such power fits on aspect ratio $x$.}
\label{fig:m_sph_dim}
\end{figure}

To be sure that these results are not affected by the particular definition of packing growth kinetics, another way of determining the parameter $d$ at saturation can be used. It bases on the dependence of the median of the number of iterations required to reach the saturation on the packing size \cite{Ciesla2017}. Namely $M_t (L) \sim L^d$, where $t$ is the random variable denoting the number of iterations needed to generate saturated packing expressed in dimensionless time units, and $d$ is the same exponent as in (\ref{eq:fl}). The dependence is shown in Figs.~\ref{fig:median} and \ref{fig:m_sph_dim}.
Exponents $d$ obtained from fitting are $1.486 \pm 0.014$, 
$1.581 \pm 0.025$, $1.852 \pm 0.023$ and 
$1.967 \pm 0.023$ for ellipses, dimers, spherocylinders and rectangles of width-to-height ratio $x=2.0$, respectively. These results confirm previous conclusions and agree with theoretical predictions for packings built of ellipses and rectangles \cite{Baule2017}. However, in order to check if the continuous character of transition of $d$ from $1.5$ to $2.0$ observed for spherocylinders and dimers of moderate anisotropy will be preserved for arbitrary large packings, the dependence of packings properties on packing size should be examined more carefully. 
\subsection{Finite size effects}
\begin{figure}[htb]
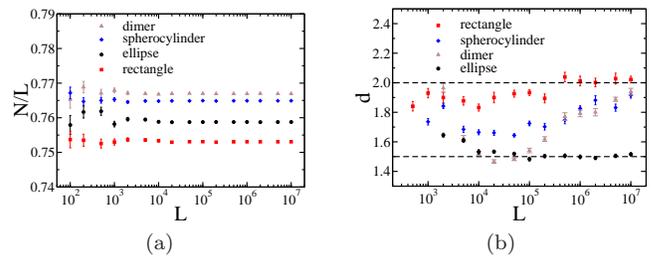

  \vspace{0.5in}
  \subfigure[]{\includegraphics[width = 0.45\columnwidth]{n_l}}
  \hspace{0.05\columnwidth}
  \subfigure[]{\includegraphics[width = 0.45\columnwidth]{d_l}}
  \caption{The dependence of the mean packing density (a) and the exponent $d$ near saturation (b) on packing size. Dots correspond to numerical data for shapes of anisotropy $x=2.0$. Dashed black lines on the right panel correspond to $d=1.5$ and $d=2$ derived analytically for shapes with analytical and non-analytical contact function \cite{Baule2017}.}
  \label{fig:q_l}
\end{figure}

As it was shown for disks, when considering packing fraction, finite size effects in RSA packings vanish along with the oscillations of the density autocorrelation function \cite{Ciesla2018}. Therefore, because the density autocorrelations vanish super-exponentially with the distance \cite{Bonnier1994}, it is not expected to observe any finite size effects for the packings studied. Fig.~\ref{fig:q_l} shows the dependence of the mean measured value of the shape density and the exponent $d$ near saturation on a packing size.
For smaller packings $L \le 10^3$, some deviations of the measured densities can be noticed but it can be rather a statistical effect due to large uncertainty of the mean density than a systematic error caused by a finite size of the packing. Parameter $d$ obtained from fitting numerical data to eq.\ref{eq:fl} varies more, but in case of packing built of ellipses and rectangles it stabilizes around $L\ge 10^5$. The situation is different for spherocylinders and dimers, where it is clear that at least for $x=2.0$ parameter $d$ increases with packing size. 

\begin{figure}[htb]
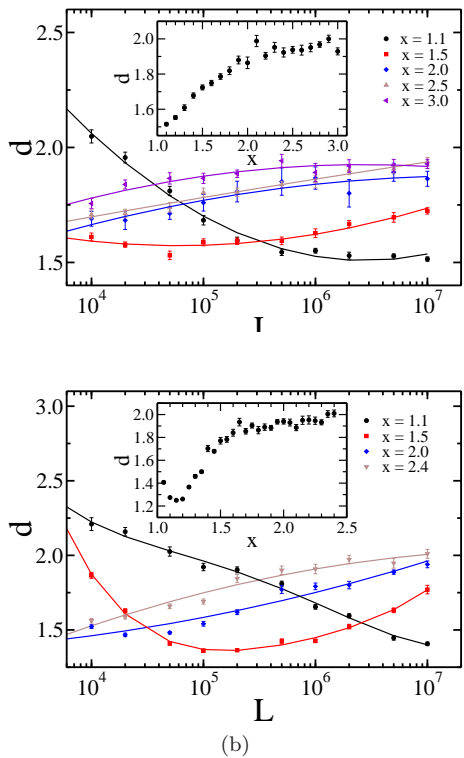

  \vspace{0.5in}
  \subfigure[]{\includegraphics[width = 0.7\columnwidth]{d_l_sph}}
  \subfigure[]{\includegraphics[width = 0.7\columnwidth]{d_l_dim}}
  \caption{Dependence of the parameter $d$ estimated from (\ref{eq:fl}) on packing size for RSA packings built of spherocylinders (a) and dimers (b) of several different anisotropies $x$. Dots are the data obtained from numerical simulation and dashed lines were drown to guide the eye. Insets show the dependence of the parameter $d$ on the anisotropy for packing size $L=10^7$.}
  \label{fig:d_l_sph_dim}
\end{figure}

Detailed analysis of this dependence for other anisotropies is shown in Fig.~\ref{fig:d_l_sph_dim}.
The stable value of $d$ is observed only for quite large anisotropies ($x=3.0$), while for smaller ones, parameter $d$ estimated from (\ref{eq:fl}) after initial decline seems to slightly grow with packing size. Moreover, for small anisotropies the rate of this growth is larger for larger $x$. The only exception to this behavior is the case of $x=1.1$, but it is possible that in this case the packing is still to small to observe any growth there. It means that the continuous transition from $d=1.5$ to $d=2.0$ can be caused by the finite size of a packing, and in the limit of infinitely large system the transition can be discontinuous. Interestingly, for dimers of small anisotropy and packing sizes $L\approx 10^5 - 10^6$, parameter $d$ is significantly below $1.5$, which is not observed for spherocylinders. 

Theoretical arguments also support the explanation of the transition as a finite size effect. In the analytical solution of the growth kinetics \cite{Baule2017}, it is shown that $d$ is determined by the analytic properties of the function $\psi(z,\alpha,\beta)=r(\alpha,z)+r(z,\beta)$ as $z$ approaches the minimum $z^*$ of $\psi$ for given $\alpha,\beta$. Here, $r$ denotes the contact function and $\alpha,\beta$ are the orientations of the particles at the left/right end of the interval of length $z$. If $\psi$ is analytic around $z^*$ as for ellipses, $d=3/2$, if it is non-analytic (piecewise-linear), $d=2$. However, depending on $\alpha,\beta$ the behaviour around $z^*$ can be either analytic or non-analytic for rectangles and spherocylinders. This implies that the asymptotic approach is governed by a superposition of power laws with $\sim t^{-2/3}$ and $\sim t^{-1/2}$ such that the true asymptotic scaling $\sim t^{-1/2}$ requires much larger $t$ to be clearly visible. Translated to the simulation of saturated packings, $L$ needs to be likewise larger for rectangles and spherocylinders than for ellipses to exhibit the correct asymptotic scaling.

\begin{figure}
  \vspace{0.5in}
  \includegraphics[width = 0.7\columnwidth]{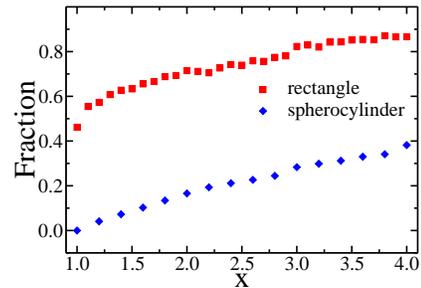}
  \caption{Fraction of non-analytic minima in $\psi(z,\alpha,\beta)$ sampled uniformly from 5000 $\alpha,\beta$ values $\in [-\pi/2,\pi/2]$ for a given aspect ratio $x$.}
  \label{fig:fraction}
\end{figure}

Following this argument, the difference in the measured $d$ values for rectangles and spherocylinders at the same value of $L$ should be due to the different frequencies at which analytic/non-analytic configurations are observed for these two shapes. To verify this argument, we have uniformly sampled orientations $\alpha,\beta$ and determined the relative frequency of non-analytic minima $z^*$ in $\psi$ (see Fig.~\ref{fig:fraction}). We see that over the range of aspect ratios considered the fraction of non-analytic minima is $>0.4$ throughout for rectangles, while it is much smaller for spherocylinders, increasing monotonically. The increase in the relative frequency for larger $x$ explains the transition observed in Fig.~\ref{fig:d_x} for fixed $L$. Moreover, Fig.~\ref{fig:fraction} also clarifies that for the same $x$, spherocylinders will need larger $L$ than rectangles to reveal the true asymptotic scaling, confirming the observations in Fig.~\ref{fig:q_l} (right).

Fig.~\ref{fig:d_l_sph_dim} likewise confirms that $d$ will generally approach the asymptotic limit for larger $L$. However, the discrepancy for $x=1.1$ is striking. It might be that for such small aspect ratios the number of configurations with non-analytic minima are simply not sufficient for the overall packing to exhibit the predicted scaling. From a theoretical perspective, this relates to the measure of such configurations in the continuous $\alpha,\beta$ range, which has not been taken into account in the analysis of \cite{Baule2017}. Extensions of the theory might thus be needed to explain the behaviour in the regime of small $x$.

Likewise, the case of dimers is special theoretically, because the function $\psi$ can exhibit continuously degenerate minima due to the non-convex shape. This case is thus not covered by the results of \cite{Baule2017} and requires further analysis. The fact that dimers behave overall very similar to spherocylinders indicates that the effect of the degeneracy might be small.
\section{Summary}
The numerical study of saturated random packings built of two-dimensional ellipses, spherocylinders, dimers, and rectangles placed on a one-dimensional line confirms the analytical results concerning the kinetics of packing growth for packing built of ellipses end rectangles \cite{Baule2017}. The first one is characterized by a power-law (\ref{eq:fl}) with the exponent $3/2$, and the second is governed by the exponent $2$. The behavior of the kinetics of packing growth for packing built of spherocylinders and dimers depends on the shape's anisotropy. For small values of the width-to-height ratio, it is the same as for ellipses, while for large values it is governed by the same exponent as for packings built of rectangles. For moderate anisotropies, and finite packing size, the continuous transition between these two regimes is observed, which can be explained for spherocylinders based on finite size arguments. In contrast to packing fraction, which quickly approaches its limiting value, parameter $d$ may vary from its value for infinitely large packing significantly even for very large systems. 
\section*{Acknowledgments}
This work was supported by grant no.\ 2016/23/B/ST3/01145 of the National Science Center, Poland. Numerical simulations were carried out with the support of the Interdisciplinary Center for Mathematical and Computational Modeling (ICM) at the University of Warsaw under grant no.\ GB76-1.
%
%
\bibliography{main}
\end{document}